\documentclass[10pt,journal,final]{IEEEtran}
\usepackage{amsmath,epsfig}
\usepackage{amssymb}
\usepackage{epic}
\usepackage{cite}
\usepackage{multirow}
\usepackage{subfigure}

\DeclareMathOperator*{\argmax}{argmax}
\newcommand{\bs}[1]{\boldsymbol{#1}}
\newcommand{\fref}[1]{Fig.~\ref{#1}}
\newcommand{\eref}[1]{(\ref{#1})}
\newcommand{\x}{\bs{x}}
\newcommand{\y}{\bs{y}}
\newcommand{\z}{\bs{z}}
\newcommand{\SetC}{\mathbb{C}}
\newcommand{\SetS}{\mathbb{S}}

 \graphicspath{{figures/}}

\begin{document}

\title{Networked Multiple Description Estimation and Compression
with Resource Scalability}

\author{Xiaolin~Wu,~\IEEEmembership{Senior Member,~IEEE,}
Xiaohan~Wang and Zhe~Wang
\thanks{Parts of this work were presented at 2006 IEEE Data
Compression Conference, Snowbird, UT, and 2007 IEEE Information Theory
Workshop, Lake Tahoe, CA.}%
\thanks{The authors are with Department of Electrical
and Computer Engineering, McMaster University, Hamilton, Ontario,
L8S 4K1, Canada (e-mail: xwu@ece.mcmaster.ca; wangx28@mcmaster.ca;nic
zwang@ece.mcmaster.ca).}}

\maketitle

\begin{abstract}

We present a joint source-channel multiple description (JSC-MD)
framework for resource-constrained network communications (e.g.,
sensor networks), in which one or many deprived encoders communicate
a Markov source against bit errors and erasure errors to many
heterogeneous decoders, some powerful and some deprived.  To keep the encoder
complexity at minimum, the source is coded into $K$ descriptions by
a simple multiple description quantizer (MDQ) with neither entropy
nor channel coding.  The code diversity of MDQ and the path
diversity of the network are exploited by decoders to correct
transmission errors and improve coding efficiency.  A key design objective
is resource scalability: powerful nodes in the network can perform JSC-MD
distributed estimation/decoding under the criteria of maximum
\emph{a posteriori} probability (MAP) or minimum mean-square error
(MMSE), while primitive nodes resort to simpler MD decoding, all
working with the same MDQ code.  The application of JSC-MD to
distributed estimation of hidden Markov models in a sensor network
is demonstrated.

The proposed JSC-MD MAP estimator is an algorithm of the longest path in a
weighted directed acyclic graph, while the JSC-MD MMSE decoder is an extension
of the well-known forward-backward algorithm to multiple descriptions.
Both algorithms simultaneously exploit the source memory, the redundancy of the
fixed-rate MDQ, and the inter-description correlations.  They outperform the
existing hard-decision MDQ decoders by large margins (up to 8dB).
For Gaussian Markov sources, the complexity of JSC-MD distributed
MAP sequence estimation can be made as low as that of typical
single description Viterbi-type algorithms.

The new JSC-MD framework also enjoys an operational advantage over
the existing MDQ decoders. It eliminates the need for multiple side
decoders to handle different combinations of the received
descriptions by unifying the treatments of all these possible cases.

\end{abstract}


\section{Introduction}

We propose a joint source-channel multiple description (JSC-MD)
framework for distributed communication and estimation of memory
sources.  The JSC-MD framework is designed to suit lossy networks populated
by resource-deprived transmitters and receivers of varied
capabilities. Such a scenario is common in sensor networks and
wireless networks.  For instance, a large number of inexpensive
sensors with no or low maintenance are deployed to monitor, assess,
and react to a large environment. On one hand these sensors have to
conserve energy to ensure a long lifespan, and on the other hand
they need to communicate with processing centers and possibly also
among themselves in volatile and adverse network conditions. The
energy budget and equipment level of the receivers vary greatly,
ranging from powerful processing centers to deprived sensors
themselves. The heterogeneity is also the norm in consumer-oriented
wireless networks.  A familiar and popular application is multimedia
streaming with mobile devices such as handsets, personal data
assistance (PDA), and notebook computers. Again battery life is a
primary concern for all mobile data transmitters, while its
criticality varies for receivers, depending on whether the receivers
are cell phones, notebooks, base stations, etc.

Conventional source and channel coding techniques may not be good
choices for networks of resource-constrained nodes, because they
make coding gains proportional to computational complexity (hence
energy consumption).  The needs for power-aware signal compression
techniques have generated renewed interests in the theory of
Slepian-Wolf and Wyner-Ziv coding, which was developed more than
thirty years ago \cite{Slepian,Wyner}.  The key insight of these
works is that statistically dependent random sources can be encoded
independently without loss of rate-distortion performance, if the
decoder has the knowledge or side information about such
dependencies.  Although originally intended for distributed source
coding, the approach of Slepian-Wolf and Wyner-Ziv coding is of
significance to resource-constrained compression in two aspects:
\begin{enumerate}
\item
communication or coordination between the encoders of the different
sources is not necessary to achieve optimal compression, even if the
sources are statistically dependent, saving the energy to
communicate between the encoders;
\item
it is possible to shift heavy computation burdens of rate-distortion
optimal coding of dependent sources from encoders to decoders.
\end{enumerate}
Such an asymmetric codec design provides an attractive signal
compression solution in situations where a large number of
resource-deprived and autonomous encoders need to communicate
multiple statistically dependent sources to one or more capable
decoders, as is the case for some hierarchical sensor networks
\cite{Akyldiz}.

Recently, many researchers have
been enthusiastically investigating practical Wyner-Ziv video coding
schemes \cite{Puri,Girod}, seeking for
energy-conserving solutions of video streaming on mobile
devices.  The motive is to perform video compression without
computationally expensive motion compensation at the encoder,
departing from the prevailing MPEG practice.  Instead, the decoder is
responsible to exploit the interframe correlations to achieve coding
efficiency.

While Wyner-Ziv coding can shift computational complexity of signal
compression from encoders to decoders, it does not address another
characteristic of modern communication networks: uneven distribution
of resources at different nodes.  As mentioned earlier, decoders can
differ greatly in power supply, bandwidth, computing capability,
response time, and other constraints.  What can be done if a decoder
has to operate under severe resource constraints as well?  Despite
the information theoretical promise of Wyner-Ziv coding, the
rate-distortion performance of distributed compression is
operationally bounded by the intrinsic complexity of the problem, or
equivalently by the energy budget.  It is well known that
optimal rate-distortion compression in centralized form is NP-hard
\cite{VQComplexity}.  We have no reason to believe that approaching
the Wyner-Ziv limit is computationally any easier.

Given the conflict between energy conservation and coding
performance, it is desirable to have a versatile signal coding and
estimation approach whose performance can be scaled to available energy,
which is the notion of resource scalability of this paper.  The
key design criterion is to keep the complexity of the encoders
(often synonymously sensors in sensor networks) at minimum, while
allowing a wide range of trade-offs between the complexity and
rate-distortion performance at decoders. Depending on the
availability of energy, bandwidth, CPU power, and other resources,
different decoders should be able to reconstruct {\em the same coded
signal(s)} on best effort basis. We emphasize that a same code
stream or a same set of code streams (in case of multiple
descriptions) of one or more sources is generated and transmitted
for an entire network.  By not generating different codes of a
source to different decoder specifications, encoders save the
energy needed to generate multiple codes.  Furthermore, this will
simplify and modularize the encoder (sensor) design to reduce the
manufacturing cost.  Ideally, a resource-scalable code should not
deny a decoder without resource constraint the possibility of
approaching the Wyner-Ziv performance limit, and at the same time it
should allow even the least capable decoder in the network to
reconstruct the signal, barring complete transmission failure.

This paper will show how resource-scalable networked signal
communication and estimation can be realized by multiple description
quantization (MDQ) at encoders and joint source-channel (JSC)
estimation at decoders.  To keep the encoder complexity at minimum,
a source is compressed by fixed rate MDQ with neither entropy nor
channel coding.  The code diversity of MDQ and the path diversity of
the network are intended to be exploited by JSC decoding to combat
transmission errors and gain coding efficiency.  Various JSC
estimation techniques will be introduced to provide solutions of
different complexities and performances, ranging from the fast and
simple hard-decision decoder to sophisticated graph theoretical
decoders.

When used for MDQ decoding, the proposed JSC-MD approach has an
added operational advantage over the current MDQ design.  It
generates an output sequence (the most probable one given the source
and channel statistics) consisting entirely of the codewords of the
central quantizer, rather than a mixture of codewords of the central
and $K$ side decoders.  As such the JSC-MD approach offers a side
benefit of unifying the treatment of the $2^K$ cases for different
subsets of received descriptions. Instead of employing $2^K-1$
decoders as required by the existing MDQ decoding process, we need
only one MDQ decoder.  This overcomes a great operational difficulty
currently associated with the MDQ decoding process.

The presentation flow of this paper is as follows.
Section~\ref{sec:formulation} formulates the JSC-MD problem.
Section~\ref{sec:MAP} constructs a weighted directed acyclic graph
to model the JSC-MD MAP estimation/decoding problem.  This graph construction
converts distributed MAP estimation into a problem of longest path in the graph,
which is polynomially solvable.  The complexity results ar derived.
Section~\ref{sec:HMM} applies the proposed JSC-MD approach to
distributed MAP estimation of hidden Markov state sequences in lossy networks.
This problem is motivated by sensor networks of heterogeneous nodes
with resource scalability requirements.  With the same MD code transmitted
over the entire network, the enpowered MD decoders can obtain exact MAP
solution using a graph theoretical algorithm, while deprived MD decoders can
obtain approximate solutions using algorithms of various complexities.
Section~\ref{Sec:MMSE} investigates the problem of distributed MMSE
decoding of MDQ.  It turns out that JSC-MD MMSE
decoding can be performed by generalizing the well-known
forward-backward algorithm to multiple descriptions.
Simulation results are reported in Section~\ref{sec:simu_results}.
Section~\ref{sec:conclusions} concludes.


\section{Problem Formulation} \label{sec:formulation}

\begin{figure}[t]
  \center
  \includegraphics[width=3.4in]{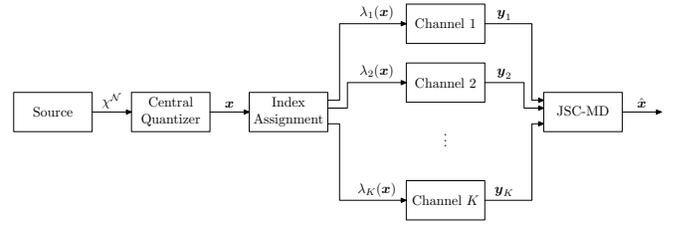}
  \caption{Block diagram of a MDQ based communication system with a
  JSC-MD decoder.}\label{Fig:system}
\end{figure}

\fref{Fig:system} schematically depicts the JSC-MD system motivated
in the introduction.  The input to the system is a
finite Markov sequence $\chi^{\cal
N}=\chi_1,\chi_2,\cdots,\chi_{\cal N}$.  A $K$-description MDQ first
maps a source symbol (if multiple description scalar quantization
(MDSQ) is used) or a block of source symbols (if multiple
description vector quantization (MDVQ) is used) to a codeword of the
central quantizer $q:\mathbb{R} \rightarrow
\SetC=\{c_1,c_2,\cdots,c_L\}$, where $L$ is the number of codecells
of the central quantizer.  Let the codebooks of the $K$ side
quantizers be $\SetC_k=\{c_{k,1},c_{k,2},\cdots,c_{k,L_k}\}$, $1
\leq k \leq K$, where $L_k \leq L$ is the number of codecells of
side quantizer $k$, $L \leq \prod_{k=1}^K L_k$.  The $K$-description
MDQ is specified by an index assignment function $\lambda_k: \SetC
\rightarrow \SetC_k$ \cite{servetto99multiple}. The redundancy
carried by the $K$ descriptions versus the single description can be
reflected by a rate $1-{\log_2 L}/{\sum_{k=1}^{K}\log_2 {L_k}}$
\cite{barros02turbo}.

Due to the expediency on the part of resource-deprived MDQ encoders,
a decoder is furnished with rich forms of statistical redundancy:
\begin{itemize}
\item the memory of the Markov source that is unexploited by suboptimal
source code;
\item
residual source redundancy for lack of entropy coding;
\item
the correlation that is intentionally introduced among the $K$
descriptions of MDQ.
\end{itemize}
The remaining question or challenge is naturally how these intra-
and inter-description redundancies can be fully exploited in a
distributed resource-constrained environment.

Let $\x=x_1 x_2 \cdots x_N \in \SetC^N$ be the output sequence of
$\chi^{\cal N}$ produced by the central quantizer, $N={\cal N}$ for
MDSQ, or $N = \iota {\cal N}$ for MDVQ with $\iota$ being the VQ
dimension.  The $K$ descriptions of MDQ, $\lambda_k(\x) \in
\SetC_k^N$, $1 \leq k \leq K$, are transmitted via $K$ noisy
diversity channels. In this work we use a quite general model for
the $K$ diversity channels.  The only requirements are that these
channels are memoryless, independent, and do not introduce phase
errors such as insertion or deletion of code symbols or bits. In the
existing literature on MDQ, only erasure errors are considered in
MDQ decoding.  Our diversity channel model accommodates bit errors
as well.  This is an important expansion because bit errors can
indeed happen in a received description in reality, particularly so
in wireless network communications.  Denote the received code
streams by $\y_k=y_{k,1}y_{k,2}\cdots y_{k,N}$, with $y_{k,n}$ being
the $n^{th}$ codeword of description $k$ that is observed by the
decoder.

Having the source and channel statistics and knowing the structure
of MDQ, the decoder can perform JSC-MD decoding of sequences $\y_k$,
$1 \leq k \leq K$, to best reconstruct $\x$.  The JSC criterion can
be maximum {\it a posteriori} probability (MAP) or minimum
mean-square error (MMSE).  For concreteness and clarity, we
formulate the JSC-MD problem for distributed MAP decoding of MDQ. As
we will see in subsequent sections, the formulation for other
distributed sequence estimation and decoding problems requires
only minor modifications.  In a departure from the
current practice of designing multiple side decoders (up to $2^K-1$
of them!), our JSC-MD system offers a single unified MDQ decoder
that operates the same way regardless what subset of the $K$
descriptions are available to the decoder.  For JSC decoding of
single description scalar quantized Markov sequences, please refer
to \cite{Park,Sayood,Subba,Wu,Zhe}.

In JSC-MD distributed MAP decoding a decoder reconstructs, given the
observed sequences $\y_k$, ($1 \leq k \leq K$, some of which may be
empty), the input sequence $\x$ such that the \emph{a posteriori}
probability $P(\x|\y_1,\y_2,\cdots,\y_K)$ is maximized. Namely, the
MAP MDQ decoder emits
\begin{equation}\label{eqn:flc_problem}
\begin{split}
\hat{\x} &= \argmax_{\x\in\SetC^N}{P(\x|\y_1,\y_2,\cdots,\y_K)}.
\end{split}
\end{equation}

Comparing the proposed JSC MDQ decoder via distributed MAP sequence
estimation with the existing symbol-by-symbol MDQ decoders, one sees
an obvious distinction.  The JSC decoder always generates codewords
of the central quantizer even when it does not have all the $K$
descriptions, while hard-decision MDQ decoders will output codewords
of side quantizers.

By Bayes' theorem we have
\begin{equation}
\begin{split} \label{eqn:flc_bayes}
&P(\x|\y_1,\y_2,\cdots,\y_K)\\
=& \frac{P(\x)P(\y_1,\y_2,\cdots,\y_K|\x)}
{P(\y_1,\y_2,\cdots,\y_K)}\\
\overset{(a)}{\propto} &P(\x)P(\y_1,\y_2,\cdots,\y_K|\x)\\
=&P(\x)P(\y_1,\y_2,\cdots,\y_K|\lambda_1(\x),\lambda_2(\x),
\cdots,\lambda_K(\x))\\
\overset{(b)}{=}&P(\x)\prod_{k=1}^{K}P(\y_k|\lambda_k(\x))\\
\overset{(c)}{=}&\prod_{n=1}^{N}\Big\{P(x_n|x_{n-1})
\prod_{k=1}^{K}P_k(y_{k,n}|\lambda_k(x_n))\Big\}.
\end{split}
\end{equation}
In the above derivation, step $(a)$ is due to the fact that $\y_1$
through $\y_K$ are fixed in the objective function for $\x \in {\cal
C}^N$; step $(b)$ is because of the mutual independency of the $K$
channels; and step $(c)$ is under the assumption that $\x$, the
output of the central quantizer, is first-order Markovian and the
channels are memoryless. This assumption is a very good approximation if
the original source sequence $\chi^{\cal N}$ before MDQ is
first-order Markovian, or a
high-order Markov sequence $\chi^{\cal N}$ is vector quantized into $K$
descriptions.

In (\ref{eqn:flc_bayes}) we also let $P(x_1|x_0)=P(x_1)$ as
convention. $P_k(\bs{b}'|\bs{b})$ is the probability of receiving a
codeword $\bs{b}=b_1 b_2 \cdots b_B$ from channel $k$ as
$\bs{b}'=b'_1 b'_2 \cdots b'_B$.  Because the channel is memoryless,
we have
\begin{equation}\label{eqn:P_k}
P_k(\bs{b}'|\bs{b}) = \prod_{i=1}^{B}P_k(b'_i|b_i).
\end{equation}

Specifically, if the $K$ diversity channels can be modeled as
memoryless error-and-erasure channels (EEC), where each bit is
either transmitted intact, or inverted, or erased (the erasure can
be treated as the substitution with a new symbol $'\$'$), then
$\bs{b} \in \{0,1\}^B, \bs{b}' \in \{0,1,\$ \}^B$ and
\begin{equation}
P_k(b'_i|b_i)=
  \begin{cases}
    p_{\phi,k}, & \text{if $b'_i=\$$}; \\
    (1-p_{\phi,k}) (1-p_{c,k}), & \text{if $b'_i=b_i$};\\
    (1-p_{\phi,k}) p_{c,k}, & \text{otherwise}
  \end{cases}
\end{equation}
where $p_{\phi,k}$ is the erasure probability and $p_{c,k}$ is the
inversion or crossover probability for channel $k$, $1 \leq k \leq
K$.

In the literature MDQ is mostly advocated as a measure against
packet erasure errors in diversity networks.  Such packet erasure
errors can be fit by the above model $P_k(b'_i|b_i)$ of binary
memoryless EEC, if a proper interleaver is used.

The proposed JSC-MD framework is also suitable for additive white
Gaussian noise (AWGN) channels.  If $b_i$ is binary phase-shift
keying (BPSK) modulated and transmitted through channel $k$ that is
AWGN, then
\begin{equation}
P_k(b'_i|b_i) = \frac{1}{\sqrt{\pi
\sigma_k}}e^{{-(b'_i-b_i)^2}/\sigma_k}
\end{equation}
where $\sigma_k$ is the noise power spectral density of channel $k$.

The prior distribution $P(x)$ and transition probability matrix
$P(x_n|x_{n-1})$ for the first-order Markov sequence $\x$ can be
determined from the source distribution and the particular MDQ in
question.

In the case of MDSQ, if the stationary probability density function
of the source is $p_s(\chi)$ and the conditional probability density
function is $p_s(\chi_n|\chi_{n-1})$, then
\begin{equation}\label{eqn:markov_model}
    P(x) = \int_{\chi:q(\chi)=x_1}p_s(\chi)d\chi
\end{equation}
and
\begin{equation}\label{eqn:markov_model2}
    P(x_n|x_{n-1}) = \frac{\iint_{\substack{\chi_1:q(\chi_1)=x_{n}\\
    \chi_2:q(\chi_2)=x_{n-1}}}
    p_s(\chi_1|\chi_2)p_s(\chi_2)d\chi_2 d\chi_1}
    {\int_{\chi:q(\chi)=x_{n-1}}p_s(\chi)d\chi}.
\end{equation}

If MDVQ is the source coder of the system, the transition
probability matrix for $P(x_n|x_{n-1})$'s can be determined
numerically either from a known close-form source distribution or
from a training set.

\section{Joint Source-Channel Multiple Description MAP Decoding}\label{sec:MAP}

In this section, we devise a graph theoretical algorithm for JSC-MD
MAP decoding algorithm.  Combining \eref{eqn:flc_problem} and
\eref{eqn:flc_bayes}, we have
\begin{equation} \begin{split}
\label{eqn:flc_new_formulation} \hat{\x} = \argmax_{\x\in\SetC^N}
\sum_{n=1}^{N}\Big\{ & \log{P(x_n|x_{n-1})}+ \\
&\sum_{k=1}^K\log{P_k(y_{k,n}|\lambda_k(x_n))}
 \Big\}.  \end{split}
\end{equation}

Because of the additivity of \eref{eqn:flc_new_formulation}, we can
structure the MAP estimation problem into the following subproblems:
\begin{equation}\label{eqn:flc_define_W}
\begin{split}
w(n,& x_n)=  \max_{\x \in \SetC^{n-1}} \sum_{i=1}^n
\Big\{\log{P(x_i|x_{i-1})}+ \\
& \sum_{k=1}^K\log{P_k(y_{k,i}|\lambda_k(x_i))}\Big\},\ x_n \in
\SetC, \ \ 1 \leq n \leq N.
\end{split}
\end{equation}
The subproblems $w(\cdot,\cdot)$ can be expressed recursively as
\begin{equation}\label{eqn:new_recursion}
\begin{split}
 &w(n,x_n)\\
=&\max_{\x \in \SetC^{n-1}}
\Big\{\sum_{i=1}^{n-1}\Big[\log{P(x_i|x_{i-1})}
+\sum_{k=1}^K \log{P_k(y_{k,i}|\lambda_k(x_i))}\Big]\\
&\phantom{\max_{\x \in \SetC^{n-1}} \Big\{} + \log{P(x_n|x_{n-1})}
+\sum_{k=1}^K \log{P_k(y_{k,n}|\lambda_k(x_n))}\Big\}\\
=&\max_{c\in\SetC}\Big\{w(n-1,c)+\log{P(x_n|c)}\Big\} \\
&+\sum_{k=1}^K \log{P_k(y_{k,n}|\lambda_k(x_n))}.
\end{split}
\end{equation}
Then, the solution of the optimization problem
(\ref{eqn:flc_problem}) is given recursively in a backward manner by
\begin{equation} \begin{split} \label{eqn:end_stage}
&\hat{x}_N = \argmax_{c \in \SetC} w(N,c).\\
&\hat{x}_{n-1} = \argmax_{c \in \SetC} \Big\{ w(n-1, c) +
\log P(\hat{x}_n|c) \Big\},\ 2 \leq n \leq N.\\
\end{split}
\end{equation}

\begin{figure}
  \centering
  \includegraphics[width=3.4in]{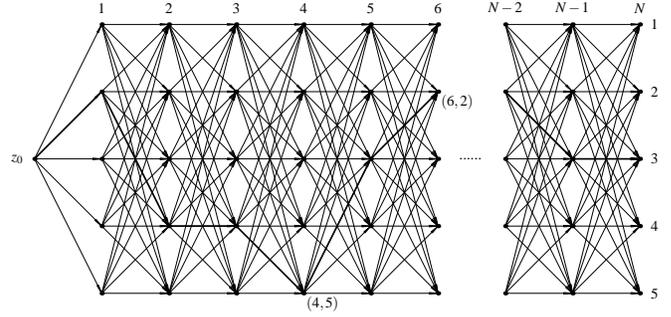}\\
  \caption{Graph $G$ constructed for the JSC-MD MAP
  decoding ($L=5$).}\label{fig:twodsq_flc}
\end{figure}

The recursion of $w(n, x_n)$ allows us to reduce the MAP estimation
problem to one of finding the longest path in a weighted directed
acyclic graph (WDAG)~\cite{Wu}, as shown in \fref{fig:twodsq_flc}.
The underlying graph $G$ has $L N+1$ vertices, which consists of $N$
stages with $L$ vertices in each stage.  Each stage corresponds to a
codeword position in $\x$. Each vertex in a stage represents a
possible codeword at the position. There is also one starting node
$z_0$, corresponding to the beginning of $\x$.

In the construction of the graph $G$, each node is associated with
a codeword $x \in \SetC$ at a sequence position $n$, $1 \leq n \leq N$,
and hence labeled by a pair $(n,x)$.  From node $(n-1,b)$ to node
$(n,a)$, $a, b \in \SetC$, there is a directed edge, whose weight is
\[\log{P(a|b)}
+\sum_{k=1}^K\log{P_k(y_{k,n}|\lambda_k(a))}.\] From the starting
node $s$ to each node $(1,a)$, there is an edge whose weight is
\[\log{P(a)} +\sum_{k=1}^K\log{P_k(y_{k,1}|\lambda_k(a))}.\]

In graph $G$, the solution of the subproblem $w(n,a)$ is the weight
of the longest path from the starting node $s$ to node $(n,a)$,
which can be calculated recursively using dynamic programming.  The
MAP decoding problem is then converted into finding the longest path
in graph $G$ from the starting node $z_0$ to nodes $(N,c),
c\in\SetC$. By tracing back step by step to the starting node $z_0$
as given in \eref{eqn:end_stage}, the MDQ decoder can reconstruct
the input sequence $\x$ to $\hat{\x}$, the optimal result defined in
\eref{eqn:flc_problem}.

Now we analyze the complexity of the proposed algorithm.  The
dynamic programming algorithm proceeds from the starting node $z_0$
to the nodes $(N,c)$, through all $L N$ nodes in $G$.  The value of
$w(n,a)$ can be evaluated in $O(L)$ time, according to
\eref{eqn:new_recursion}. The quantities $\log{P(a|b)}$ and
$\log{P_k(y_{k,n}|\lambda_k(a))}$ can be precomputed and stored in
lookup tables so that they will be available to the dynamic
programming algorithm in $O(1)$ time. Hence the term $\sum_{k=1}^K
\log{P_k(y_{k,n}|\lambda_k(a))}$ in \eref{eqn:new_recursion} can be
computed in $O(K)$ time.  Therefore, the total time complexity of the
dynamic programming algorithm is $O(L^2 N K)$.  The reconstruction
of the input sequence takes only $O(N)$ time, given that the
selections in \eref{eqn:end_stage} (and in \eref{eqn:new_recursion}
as well) are recorded, which results in a space complexity of $O(L
N)$.

In~\cite{Wu} we proposed a monotonicity-based fast algorithm for the
problem of MAP estimation of Markov sequences coded by a single
description quantizer, which converts the longest path problem to
one of matrix search.  For Gaussian Markov sequences the matrix can
be shown to be totally monotone, and the search can be done in
lower complexity.  The same algorithm technique can be generalized
to multiple descriptions and reduce the complexity of JSC-MD MAP
decoding.  In the appendix we prove that distributed MAP decoding
of $K$-description scalar quantizer can be completed in
$O(L N K)$ time for Gaussian Markov sequences.
The linear dependency of the MAP MDSQ decoding algorithm in the sequence
length $N$ and source codebook size $L$ makes it comparable to the
complexity of typical Vertibi-type decoders for single description.

\section{Distributed Multiple-Description Estimation of Hidden Markov Sequences}
\label{sec:HMM}

In this section we apply the proposed JSC-MD MAP estimation technique
to solve the problem of hidden Markov sequence estimation in a
resource-constrained network.  For single description hidden Markov sequence
estimation is an extensively-studied problem with many applications
\cite{Rabiner}.  As a case study, consider a sensor network
in an inaccessible area to monitor the local weather
system for years with no or little maintenance. Our objective is to
remotely estimate the time sequence of weather patterns: sunny,
rainy, cloudy and so on.  To this end the sensors collect
real-valued data vector: temperature, pressure, moisture, wind
speed, etc., and communicate them to processing nodes of various
means in the network.  Some are well-equipped and easily-maintained
processing centers, while others need to run autonomously on limited
power supply and react to certain weather conditions on their own
rather than being instructed by the central control.

\subsection{Problem Formulation}

Our task is to estimate the state sequence of a hidden Markov model (HMM),
which is, in our example, the time sequence of weather patterns
that are not directly observable
by processing nodes in the sensor network.  Let the state
space of the HMM be $\SetS = \{s_1, s_2, \cdots, s_M\}$, being
sunny, rain, and so on. For state transition from
$s_i$ to $s_j$ (weather change) of Markov state transition
probability $P_S(s_j | s_i)$, the HMM output to be observed by the
sensors is a real-valued random vector $x \in \mathbb{R}^d$ (temperature,
pressure, moisture, wind speed, etc.) with
probability $P_O(x | s_i,s_j)$.  The observations need to be
communicated to data processing centers at a low bit rate against
channel noise and losses.  To maximize their operational lifetime the sensors
have to do without sophisticated source coding and forgo channel
coding altogether.  A viable solution under such stringent conditions is
to produce and transmit $K \geq 2$ descriptions of $x$
in fixed length code without entropy coding.  There are many ways for
inexpensive and deprived encoders (sensors) to code $x$ into multiple
descriptions in collaboration.  One possibility is the use of
multiple description lattice vector quantizer (MDLVQ) \cite{servetto99multiple,Vaishampayan_flc,Huang2006}.

\begin{figure}[thb]
\centering
  \includegraphics[width=2.8in]{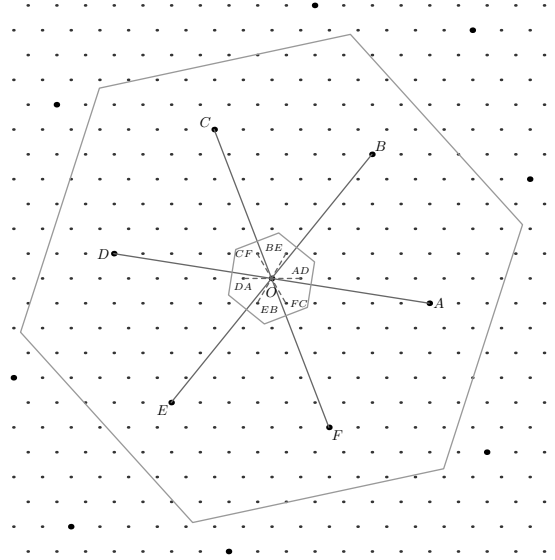}\\
  \caption[MDLVQ index assignment for $A_2$ lattice, $K=2$]
  {MDLVQ index assignment for $A_2$ lattice, $K=2$.  Points
  of $\SetC$ and $\SetC_s$
  are marked by $\cdot$ and $\bullet$, respectively.}
  \label{fig:N31IA}
\end{figure}

Among known multiple description vector codes, MDLVQ is arguably the
most resource-conserving with a very simple implementation.  A
$K$-description MDLVQ uses a fine lattice in $\mathbb{R}^d$ as its
central quantizer codebook $\SetC$ and an accompanying coarse
lattice $\SetC_s$ in $\mathbb{R}^d$ as its side quantizer codebooks
$\SetC_k,\ 1\leq k\leq K$.  Therefore we have
$\SetC_1=\SetC_2=\cdots=\SetC_K=\SetC_s$, and typically $\SetC_s
\subset \SetC$. An MDLVQ index assignment is depicted in
Fig.~\ref{fig:N31IA} for $K=2$.  Each fine lattice point in $\SetC$ is labeled
by a unique ordered pair of coarse lattice points in $\SetC_s$.

Upon observing an HMM output sequence $\chi^{N}$, the central
quantizer first quantizes $\chi^N$ to a sequence of the nearest fine
lattice points $\x = q(\chi^N)$.  Then the MDLVQ encoder generates
$K$ description sequences of $\x$: $\lambda_k(\x)$ and transmits
them through $K$ diversity channels (or diversity paths in the
network).

A decoder can reconstruct $\lambda_k(x_n)$ to $x_n$ with the inverse
labeling function $\lambda^{-1}$, if all $K$ descriptions are
received. In the event that only a subset $\Psi$ of the $K$
descriptions are received, the decoder reconstructs $x_n$ to the
average of the received coarse descriptions:
\begin{equation}\label{hard-decision}
\hat{x}_n = \frac{1}{|\Psi|} \sum_{k \in \Psi} \lambda_k(x_n)
\end{equation}
where $|\cdot|$ is the cardinality of a set.  This is the simplest
MDLVQ decoder possible, which is also asymptotically optimal for
$K=2$ \cite{SVS2001}.

\subsection{Distributed MAP Sequence Estimation}

Let $\y_k = y_{k,1} y_{k,2} \cdots y_{k,N}$ be the received sequence
from channel $k$, $1 \leq k \leq K$.  Our task is to estimate the
hidden state sequence $\z = z_1 z_2 \cdots z_N \in \SetS^N$ of
weather patterns, given the $K$ noisy time sequences of atmosphere
attributes produced by the HMM: $\y_1$, $\y_2$, ..., $\y_K$.  With
the resource-scalability in mind, we take an approach of MAP
estimation:
\begin{equation}\label{eqn:hmm_map}
\begin{split}
\hat{\z}, \hat{\x} = \argmax_{\z\in\SetS^N, \x \in \SetC^N} {P(\x,
\z|\y_1,\y_2,\cdots,\y_K)}.
\end{split}
\end{equation}

Analogously to (\ref{eqn:flc_bayes}) we use Bayes' theorem and the
independence of the $K$ memoryless channels to obtain
\begin{equation}\label{eqn:hmm_bayes}
\begin{split}
&P(\x, \z|\y_1,\y_2,\cdots,\y_K)\\
\propto & P(\z) P(\x|\z) P(\y_1,\y_2,\cdots,\y_K|\x, \z)\\
=&\prod_{n=1}^N \Big\{ P_S(z_n|z_{n-1})P_O(x_n|z_n,z_{n-1})
 P(\y_1,\y_2,\cdots,\y_K|\x) \Big\} \\
=&\prod_{n=1}^N \Big\{ P_S(z_n|z_{n-1}) P_O(x_n|z_n,z_{n-1})
\prod_{k=1}^K P_k(y_{k,n}|\lambda_k(x_n)) \Big\}.
\end{split}
\end{equation}

We can also devise a graph theoretical algorithm to solve the
sequence estimation problem of \eref{eqn:hmm_bayes}. Combining
\eref{eqn:hmm_map} and \eref{eqn:hmm_bayes} and taking logarithm, we
have
\begin{equation} \begin{split}
\label{eqn:hhm_new_formulation} \hat{\z}, \hat{\x} &= \argmax_{\z
\in \SetS^N, \x \in \SetC^N} \sum_{n=1}^N \Big\{ \log
P_S(z_n|z_{n-1}) + \\
&\log P_O(x_n|z_n,z_{n-1}) + \sum_{k=1}^K \log
P_k(y_{k,n}|\lambda_k(x_n)) \Big\}.
\end{split}
\end{equation}
Then, the MAP estimate of the sequence of hidden Markov states is
given by
\begin{equation} \begin{split}
\label{eqn:hmm_states} \hat{\z} = \argmax_{\z \in \SetS^N}
\sum_{n=1}^N \Big\{ \log P_S(z_n|z_{n-1}) + \xi_n(\z) \Big\}
\end{split}
\end{equation}
where
\begin{equation}\nonumber
\xi_n(\z) = \max_{x \in \Lambda} \Big\{\log P_O(x|z_n,z_{n-1}) +
\sum_{k=1}^K \log P_k(y_{k,n}|\lambda_k(x))\Big\}.
\end{equation}

Using the same technique as used in \eref{eqn:new_recursion}, we
structure the above optimization problem into a nested set of
subproblems:
\begin{equation}\label{eqn:hmm_define_W}
\begin{split}
w(n,z_n)=\max_{\z \in \SetS^{n-1}} \sum_{i=1}^n \Big\{ \log
P_S(z_i|z_{i-1}) +  \xi_i(\z) \Big\} \\
z_n \in \SetS, \ \ 1 \leq n \leq N
\end{split}
\end{equation}
which can be expressed recursively by
\begin{equation}\label{eqn:HMM_new_recursion}
w(n,z_n) = \max_{s\in\SetS} \Big\{ w(n-1,s)+\log P_S(z_n|s) +
\xi(z_n,s) \Big\}
\end{equation}
where
\begin{equation}\nonumber
\xi(z_n,s) = \max_{x \in \SetC} \Big\{ \log P_O(x|z_n,s) +
\sum_{k=1}^{K} \log P_k(y_{k,n}|\lambda_k(x)) \Big\}.
\end{equation}

This recursion form also enables us to solve the sequence estimation
problem of \eref{eqn:hmm_bayes} by finding the longest path in a
WDAG.  The WDAG $G$ contains $M N+1$ vertices: a starting node $z_0$
and $N$ stages with $M$ vertices in each stage.  Each stage
corresponds to a position in time sequence $\z$.  Each vertex in a
stage represents a possible HMM state at the position. The starting
node $z_0$ corresponds to the beginning of the sequence $\z$.  From
node $(n-1,a)$ to node $(n,b)$, $a, b \in \SetS$, there is a
directed edge with weight:
\[ \log P_S(b|a) + \xi(b,a). \]
>From the starting node $z_0$ to each node $(1,a)$, there is an edge
whose weight is
\[ \log P_S(a) + \max_{x\in \SetC} \Big\{\log P_O(x|a) +
\sum_{k=1}^K\log{P_k(y_{k,1}|\lambda_k(x))}\Big\}. \]

In graph $G$, the solution of the subproblem $w(n,s)$ is the weight
of the longest path from the starting node $z_0$ to node $(n,s)$,
which can be calculated recursively using dynamic programming.  The
distributed MAP estimation problem is then converted into finding
the longest path in graph $G$ from the starting node $z_0$ to nodes
$(N,s), s\in\SetS$. Tracing back step by step to the starting node
$z_0$ generates the optimally estimated HMM state sequence
$\hat{\z}$.

To analyze the complexity of the proposed algorithm, we notice that
the dynamic programming algorithm proceeds through all $M N$ nodes
in $G$.  The value of $w(n,s)$ can be evaluated in $O(M)$ time,
according to \eref{eqn:HMM_new_recursion}. The quantities
$\log{P_S(b|a)}$ and $\log P_O(x|z_n,s)$ can be precomputed and
stored in lookup tables so that they will be available in the
dynamic programming process in $O(1)$ time.  The term $\xi(b,a)$ can
be computed in $O(KL)$ time. Therefore, the total time complexity of
this algorithm is $O(M^2 N K L)$.  The space complexity is $O(M N)$.

\subsection{Resource Scalability}

If a network node is not bounded by energy and computing resources,
it can use the relatively expensive MAP algorithm that taps all
available redundancies to obtain the best estimate of HMM state
sequence, knowing the statistics of HMM and underlying noisy
diversity channels.  This JSC-MD framework can be used as an
asymmetric codec in the Wyner-Ziv spirit, which stripes the encoders
to the bone while enpowering the decoders.  More importantly, it
also offers a resource-scalability. If a node in the sensor network
needs to estimate $\z$ but is severely limited in resources, it can
still do so using the same MDLVQ code, albeit probably at a lesser
estimation accuracy. The simplest hence most resource-conserving
approximate solution is to first perform a hard-decision MDLVQ
decoding of received descriptions $y_{k,n}$'s to $\hat{x}_n$ using
(\ref{hard-decision}), and then estimate $z_n$ to be
\begin{equation}\label{eqn:cheapest}
\hat{z}_n = \argmax_{z \in \SetS} P_S(z)P_O(\hat{x}_n|z).
\end{equation}
The hard-decision MDLVQ decoding takes only $O(K)$ operations. Also,
in the above approximation, we replace $P_S(z_n|z_{n-1})$ by
$P_S(z_n)$ and $P_O(\hat{x}_n|z_n,z_{n-1})$ by $P_O(\hat{x}_n|z_n)$
in (\ref{eqn:hmm_bayes}).  This is to minimize the resource
requirement for estimating $\z$ by ignoring the source memory.
Consequently, the total time complexity of the fast algorithm
reduces to $O(N(K+M))$, as opposed to $O(M^2 N K L)$ for the full
fledged MAP sequence estimation algorithm.  The space requirement
drops even more drastically to $O(M+K)$ from $O(MN)$.

Between the exact $O(M^2 N K L)$ graph theoretical algorithm and the
least expensive $O(N(K+M))$ algorithm, many trade-offs can be made
between the resource level and performance of the decoder.  For
instance, if only $P_S(z_n|z_{n-1})$ is replaced by $P_S(z_n)$ in
(\ref{eqn:hmm_bayes}), another possible JSC-MD estimation emerges:
\begin{equation}
\hat{z}_n = \argmax_{z \in \SetS} P_S(z) P_O \Big( \max_{x \in
\SetC} \prod_{k=1}^K P_k\big(y_{k,n}\big|\lambda_k(x)\big) \Big| z
\Big).
\end{equation}
This leads to an $O(N(KL+M))$ HMM state sequence estimation
algorithm.  The algorithm is slightly more expensive than the one
based on (\ref{eqn:cheapest}) but offers better performance because
the MDLVQ decoding is done with the knowledge of channel statistics.

\subsection{Application in Distributed Speech Recognition}
Given the success of HMM in speech recognition \cite{Rabiner}, we
envision the potential use of the JSC-MD estimation technique for
remote speech recognition.  For instance, the cell phones transmit
quantized speech signals via diversity channels to processing
centers and the recognized texts are sent back or forward to other
destinations.  This will offer mobile users speech recognition
functionality without requiring heavy computing power on handsets
and fast draining batteries.  Also,
the network speech recognizer can prompt a user to repeat in case of
difficulties, the user's revocalization can be used as extra
descriptions to improve the JSC-MD estimation performance.

\section{Resource-scalable JSC-MD MMSE Decoding} \label{Sec:MMSE}

The JSC-MD distributed MAP estimation problem discussed above is to
track the discrete states of a hidden Markov model.  Likewise, the
cost function (\ref{eqn:flc_problem}) for distributed MAP decoding
of MDQ requires the output symbols to be discrete codewords of the
central quantizer. This may be desirable or even necessary, if the
quantizer codewords communicated correspond to discrete states of
semantic meanings, such as in some recognition and classification
applications.  But in network communication of a continuous signal
$\chi^{\cal N}=\chi_1,\chi_2,\cdots,\chi_{\cal N}$, the JSC-MD
output can be real valued.  In this case a JSC-MD distributed MMSE
decoding scheme of resource scalability is preferred, which is the
topic of this section.


The goal of the JSC-MD MMSE decoding is to reconstruct $\chi_n$ as
\begin{equation}
\begin{split}
&E(\chi_n | \y_1; \y_2; \cdots; \y_K)\\ = &\sum_{l=1}^L P(x_n=l |
\y_1; \y_2; \cdots; \y_K) \frac{\int_{\chi \in V_l} \chi p(\chi)
d\chi}{\int_{\chi \in V_l} p(\chi) d\chi}
\end{split}
\end{equation}
where $V_l$ is cell $l$ of the central quantizer. Hence we need to
estimate the \emph{a posteriori} probability $P(x_n | \y_1, \y_2,
\cdots, \y_K)$.  Equivalently, we estimate
\begin{equation}
P(x_n=l, \y_1, \y_2, \cdots, \y_K), \ \ l \in {\SetC}.
\end{equation}

We can solve the above estimation problem for the JSC-MD distributed
MMSE decoding by extending the well-known BCJR (forward-backward)
algorithm \cite{Bahl} to multiple observation sequences.  For
notational convenience let $\y_k^{a \sim b}$, $a < b$, be the
consecutive subsequence $y_{k,a}, y_{k,a+1}, \cdots, y_{k,b}$ of an
observation sequence $\y_k$.  Define
\begin{equation}
\begin{split}
&\alpha_n(l) = P(x_n=l, \y_1^{1 \sim n}, \y_2^{1 \sim n}, \cdots,
\y_K^{1 \sim n})\\
&\beta_n(l) = P(\y_1^{n+1 \sim N}, \y_2^{n+1 \sim N}, \cdots,
\y_K^{n+1 \sim N} | x_n = l)\\
&\gamma_n(l',l) = P(x_n=l, y_{1,n}, y_{2,n}, \cdots, y_{K,n} |
x_{n-1} = l').
\end{split}
\end{equation}
Then we have
\begin{equation}
\begin{split}\label{Eqn:prob}
\lefteqn{P(x_n=l, \y_1, \y_2, \cdots, \y_K)} \\
 =& P( x_n=l,
\y_1^{1 \sim n}, y_2^{1 \sim n}, \cdots, y_K^{1 \sim n} ) \\
&\cdot
 P(\y_1^{n+1 \sim N}, \y_2^{n+1 \sim N}, \cdots,
 \y_K^{n+1 \sim N}|x_n=l)  \\ =& \alpha_n(l)\cdot \beta_n(l).
\end{split}
\end{equation}
The last step is due to the fact that $\y_k^{1 \sim n}$ and
$\y_k^{n+1 \sim N}$ are independent given $x_n$, and that $\y_k$ and
$\y_j$ are independent for $k \not = j$.  The terms $\alpha_n(l)$
and $\beta_n(l)$ can be recursively computed by
\begin{equation}\begin{split}\label{Eqn:alpha}
&\alpha_n(l)\\ =& \sum_{l'=0}^{L-1} P(x_{n-1}=l', x_n=l, \y_1^{1
\sim
n}, \y_2^{1 \sim n}, \cdots, \y_K^{1 \sim n} )  \\
=& \sum_{l'=0}^{L-1} \Big\{P(x_{n-1}=l', \y_1^{1 \sim n-1}, \y_2^{1
\sim n-1}, \cdots, \y_K^{1 \sim n-1} )\\
&\phantom{\sum_{l'=0}^{L-1}\{}\cdot P(x_n=l, y_{1,n},
y_{2,n}, \cdots, y_{K,n} | x_{n-1} = l' ) \Big\} \\
=& \sum_{l'=0}^{L-1} \alpha_{n-1}(l') \cdot \gamma_n(l',l).
\end{split}\end{equation}
and
\begin{equation}\begin{split}\label{Eqn:beta}
&\beta_n(l)\\ =& \sum_{l'=0}^{L-1} P(x_{n+1}=l', \y_1^{n+1 \sim N},
\y_2^{n+1 \sim N}, \cdots, \y_K^{n+1 \sim N} | x_n = l) \\
=& \sum_{l'=0}^{L-1} \Big\{ P(x_{n+1}=l', y_{1,n+1},
y_{2,n+1},\cdots,
y_{K,n+1} | x_n = l ) \\
&\phantom{\sum_{l'=0}^{L-1}\{}\cdot P(\y_1^{n+2 \sim N}, \y_2^{n+2
\sim N}, \cdots, \y_K^{n+2 \sim N} | x_{n+1} = l') \Big\} \\
=& \sum_{l'=0}^{L-1} \gamma_{n+1}(l,l') \cdot \beta_{n+1}(l').
\end{split}
\end{equation}


By definition the term $\gamma_n(\cdot,\cdot)$ can be computed by
\begin{equation}\begin{split}\label{Eqn:gamma}
&\gamma_n(l',l)\\ =& P(x_n=l, y_{1,n}, y_{2,n}, \cdots, y_{K,n} |
x_{n-1} = l') \\
=& P(x_n=l|x_{n-1}=l') \cdot Pr(y_{1,n}, y_{2,n},
\cdots, y_{K,n} | x_n=l)  \\
=& P(x_n=l|x_{n-1}=l') \cdot \Pi_{k=1}^K P_k (y_{k,n} | \lambda_k(l)
).
\end{split}
\end{equation}
If the input sequence $\x$ is i.i.d.\, the above is reduced to
\begin{equation}\begin{split} \label{Eqn:iidMMSE}
&P(x_n=l, \y_1, \y_2, \cdots, \y_K)\\ = &P(x_n=l, y_{1,n},
y_{2,n}, \cdots, y_{K,n}) \\
= &P(x_n=l) \cdot P(y_{1,n}, y_{2,n}, \cdots, y_{K,n}|x_n=l)
\\
= &P(x_n=l) \cdot \Pi_{k=1}^K P_k (y_{k,n}| \lambda_k(l) ), \ \ 1
\leq n \leq N.
\end{split}\end{equation}
This is also the scheme for hard-decision MDQ MMSE decoding in midst
of inversion and erasure errors.

Now we analyze the complexity of the proposed JSC-MD MMSE algorithm.
For each $n$, $1 \leq n \leq N$, we need to calculate the value of
$\alpha_n(l)$, $\beta_n(l)$ and $\gamma_n(l',l)$.  As explained in
the complexity analysis of JSC-MD MAP algorithm, the term
$\sum_{k=1}^K \log{P_k(y_{k,n}|\lambda_k(a))}$ in (\ref{Eqn:gamma})
can be computed in $O(K)$ time.  Thus, the matrix $\gamma_n(l',l)$,
$(l,l') \in {\SetC}^2$, can be computed in $O(L^2 K)$ time. The
value of $\alpha_n(l)$ and $\beta_n(l)$ can be computed in $O(L)$
time. Therefore, the total complexity is $O(L^2 K N)$, which has the
same order with the complexity of JSC-MD MAP algorithm as derived in
Section~\ref{sec:MAP}.

If sequence $\x$ is i.i.d.\ the complexity of JSC-MD MMSE decoding
is reduced to $O(LKN)$ as exhibited by (\ref{Eqn:iidMMSE}).  For
memoryless sources, MMSE sequence estimation is degenerated to MMSE
symbol-by-symbol decoding.  Even if $\x$ is not memoryless, in
consideration of resource scalability, (\ref{Eqn:iidMMSE}) can still
be used as a less demanding alternative for network nodes not having
sufficient resources to perform full-fledged JSC-MD MMSE decoding.
The approximation is good if the source memory is weak. To the
extreme, the weakest network nodes of severe source constraints can
always resort to a hard-decision MD decoding (e.g., using the MD
decoder (\ref{hard-decision})), which takes only $O(KN)$ time to
decode a multiple-description coded sequence $\x$ of length $N$. The
important point is that all three decoders of complexities ranging
from $O(L^2 K N)$ to $O(KN)$ operate on the same MD code streams
distributed in the network.  The reader can continue to the next section
for further discussions on the issue of resource scalability.

\section{Simulation Results} \label{sec:simu_results}

\begin{figure}[t]
  \center
  \label{fig:sub:b}
  \includegraphics[width=1.2in]{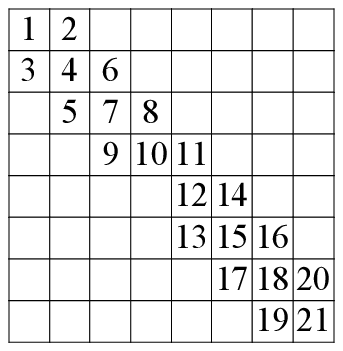}
  \caption{The index assignment for two two-description scalar
  quantizer as proposed by \cite{Vaishampayan_flc}.}
  \label{Fig:ia4cc}
\end{figure}

The proposed resource-aware JSC-MD distributed MAP and MMSE decoding
algorithms are implemented and evaluated via simulations.  The
simulation inputs are first-order, zero-mean, unit-variance Gaussian
Markov sequences of different correlation coefficient $\rho$. A
fixed-rate two-description scalar quantizer (2DSQ) proposed in
\cite{Vaishampayan_flc} is used as the encoder in our simulations.
The 2DSQ is uniform and is specified by the index assignment matrix
shown in \fref{Fig:ia4cc}.  The central quantizer has $L=21$
codecells and the two side quantizers each has $L_1 = L_2 = 8$
codecells.  For each description $k$, $k=1,2$, the codeword index
$\lambda_k(x)$ is transmitted in fixed length code of three bits.


The channels are simulated to be error-and-erasure channels with
identical erasure probability $p_\phi$ and inversion probability
$p_c$ varying.  We report and discuss below the simulation results
for different combinations of $p_c$, $p_\phi$ and $\rho$.

\begin{figure}[t]
  \center
  \includegraphics[width=3.2in]{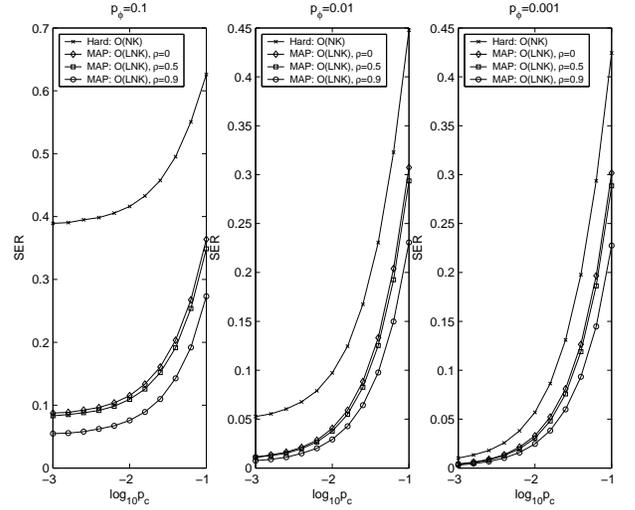}
  \caption{Symbol error rates of JSC-MD distributed MAP decoder and
  MDQ hard-decision decoder with $\rho = 0, 0.5, 0.9$.}
  \label{Fig:map}
\end{figure}

First, we evaluate the performance of the JSC-MD distributed MAP
decoder.  The performance measure is symbol error rate (SER), which
is the probability that a symbol of the input Markov sequence is
incorrectly decoded.  Since the input source is Gaussian Markov, the
$O(LKN)$ MAP algorithm of Section \ref{sec:MAP} can be used by the
resource-rich network nodes to obtain the optimal estimation.
However, resource-deprived network nodes can also decode whatever
received description(s) of the same 2DSQ code, using a simple
energy-conserving $O(KN)$ hard-decision MDQ decoder.  The simulation
results are plotted in Fig.~\ref{Fig:map}.  Over all values of
$\rho$, $p_c$ and $p_\phi$, the JSC-MD MAP decoder outperforms the
hard-decision MDQ decoder.  As expected, the performance gap between
the two decoders increases as the amount of memory in the Markov
source ($\rho$) increases.  This is because the hard-decision MDQ
decoder cannot benefit from the residual source redundancy left by
the suboptimal primitive 2DSQ encoder.

\begin{figure}[t]
  \center
  \includegraphics[width=3.2in]{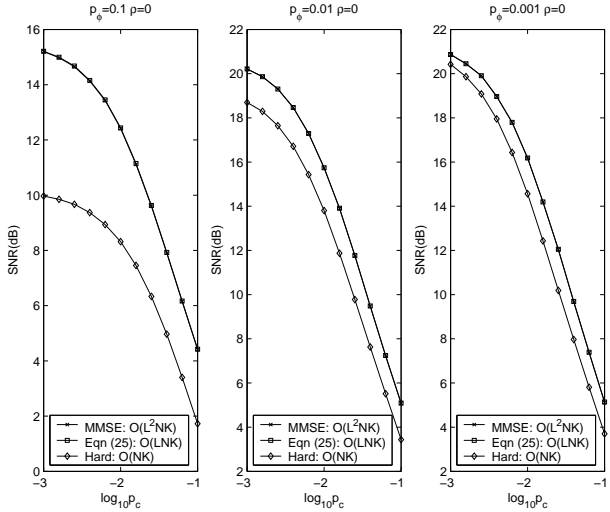}
  \caption{SNR performances of different MDQ decoders ($\rho=0$).}
  \label{Fig:MMSE0}
\end{figure}

\begin{figure}[t]
  \center
  \includegraphics[width=3.2in]{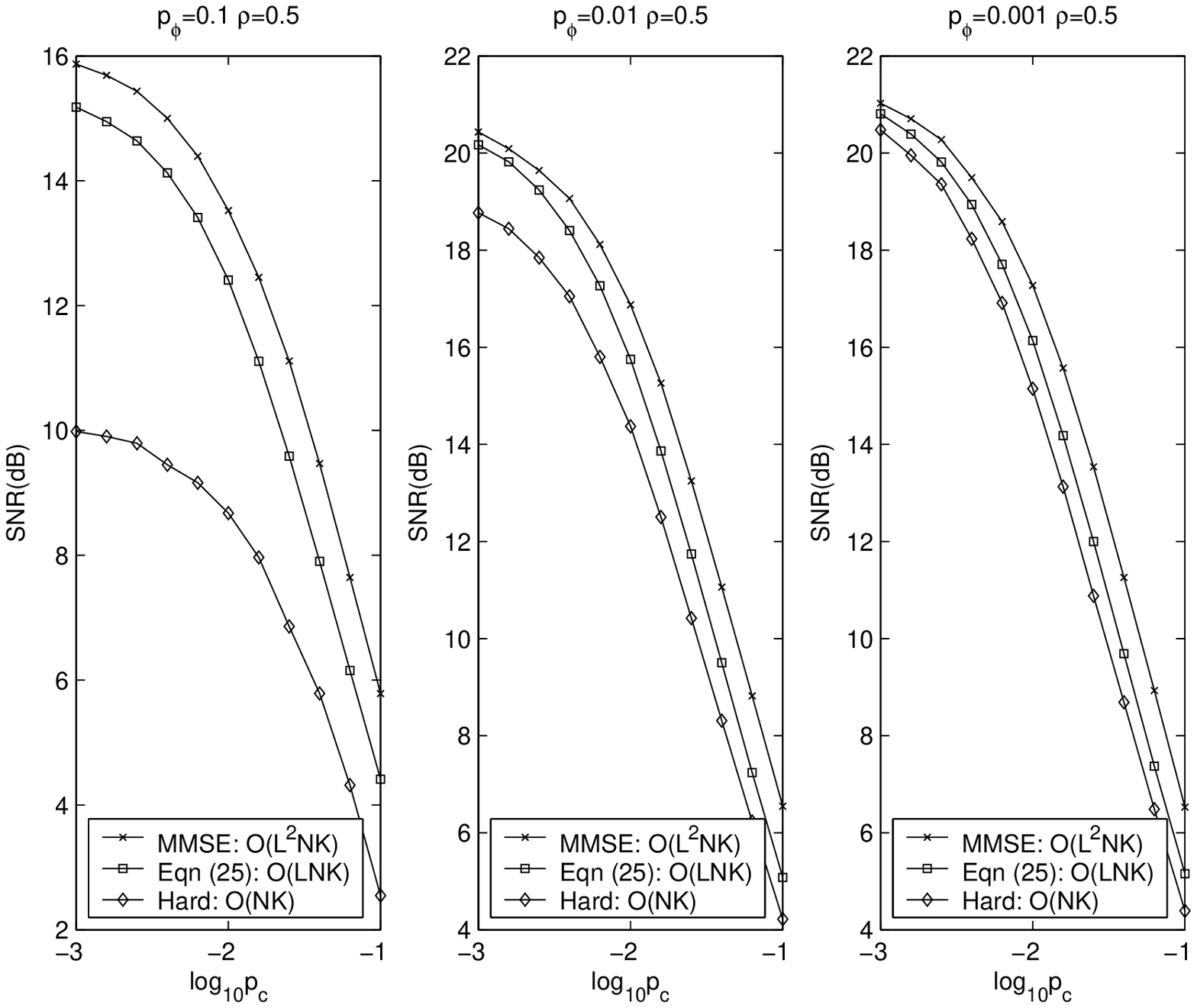}
  \caption{SNR performances of different MDQ decoders ($\rho=0.5$).}
  \label{Fig:MMSE5}
\end{figure}

\begin{figure}[t]
  \center
  \includegraphics[width=3.2in]{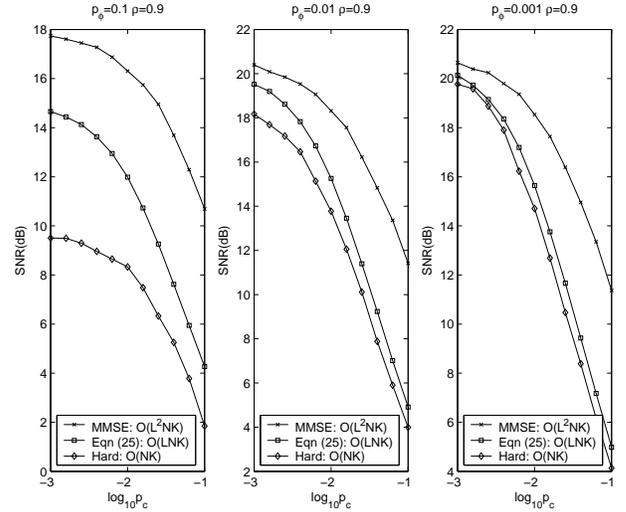}
  \caption{SNR performances of different MDQ decoders ($\rho=0.9$).}
  \label{Fig:MMSE9}
\end{figure}

In the case of JSC-MD distributed MMSE decoding, we evaluate three
decoders of different complexities (hence different resource
requirements): the exact $O(L^2 K N)$ algorithm derived in Section
\ref{Sec:MMSE}, the simplified $O(LKN)$ algorithm given in
\eref{Eqn:iidMMSE}, and the conventional $O(KN)$ hard-decision MDQ
decoder. The performance measure for MMSE decoding is naturally the
signal-to-noise ratio (SNR).  The simulation results are plotted in
Fig.~\ref{Fig:MMSE0}-\ref{Fig:MMSE9}, with the correlation
coefficient being 0, 0.5 and 0.9 respectively.  The trade-offs
between the complexity and performance of a decoder can be clearly
seen in these figures.  Given $\rho$, $p_c$, $p_\phi$, the SNR
increases as the decoder complexity increases.  The JSC-MD MMSE
decoder achieves the highest SNR, because it utilizes both inter-
and intra-description correlations.  The performance of the
algorithm given in \eref{Eqn:iidMMSE} is in the middle, which is
$O(L)$ faster than the full-fledged JSC-MD MMSE decoder but $O(L)$
slower than the hard-decision MDQ decoder.  This decoder reduces
complexity or energy requirement by making use of the
inter-description correlation only.  The hard-decision MDQ decoder
is the simplest and fastest.  However, it ignores both intra- and
inter-description correlations and has the lowest SNR.  As in the
MAP case, the performance gap between the first two MMSE decoders
increases as the intra-description redundancy ($\rho$) increases.
When $\rho=0$, the first two algorithms become the same.


Under both MAP and MMSE criteria, the performance gap between
different algorithms increases as the erasure error probability
$p_\phi$ increases, indicating that the JSC-MD distributed decoder
can make a better use of inter-description correlation in the event
of packet loss.  As the erasure error probability increases in the
network, the proposed JSC-MD decoder enjoys up to 8 dB gain over the
hard-decision MD decoders.

Finally, we point out that even when source memory is weak (see the
curves for $\rho=0$), the JSC-MD distributed decoders still have an
advantage over the hard-decision MDQ decoders that cannot handle the
bit errors within a received description effectively.

\section{Conclusions}\label{sec:conclusions}

We propose a joint source-channel multiple description approach to
resource-scalable network communications.  The encoder complexity is
kept to the minimum by fixed rate multiple description quantization.
The resulting MD code streams are distributed in the network and can
be reconstructed to different qualities depending on the resource
levels of receiver nodes.  Algorithms for distributed MAP and MMSE
sequence estimation are developed, and they exploit intra- and
inter-description redundancies jointly to correct both bit errors
and erasure errors. The new algorithms outperform the existing
hard-decision MDQ decoders by large margins (up to 8dB).  If the
source is Gaussian Markov, the complexity of the JSC-MD distributed
MAP estimation algorithm is $O(L N K)$, which is the same as the
classic Viterbi algorithm for single description.

Operationally, the new MDQ decoding technique unifies the treatments
of different subsets of descriptions available at a decoder,
overcoming the difficulty of having a large number of side decoders
that hinders the design of a good hard-decision MDQ decoder.

\appendix[Complexity Reduction of JSC-MD Problem]\label{apd:monge}

The complexity of the JSC-MD MAP decoding problem in Section
\ref{sec:MAP} can be reduced because it has a strong monotonicity
property, if the source is Gaussian Markovian and is coded
by multiple description scalar quantizer (MDSQ).  To show this we need
to convert the recursion formula in Section \ref{sec:MAP} into a matrix
search form \cite{Wu}.  We rewrite \eref{eqn:new_recursion} as
\begin{equation}\label{eqn:complexity_reduction}
\begin{split}
w(n,a)=\max_{b\in\SetC}\Bigl\{ & w(n-1,b)+\log{P(a|b)}\\
&+\sum_{k=1}^K \log{P_k(y_{k,n}|\lambda_k(a))}\Bigr\}.
\end{split}
\end{equation}
Then for each $1\leq n \leq N$, we define an $L\times L$ matrix
$A_n$ such that
\begin{equation}\label{eqn:matrix_definition}
\begin{split}
A_n(a,b) = w(n-1,b)+\log{P(a|b)}+ \sum_{k=1}^K
\log{P_k(y_{k,n}|\lambda_k(a))}.
\end{split}
\end{equation}
Now one can see that the computation task for JSC-MD MAP decoding
is to find the row maxima of matrix $A_n$.

A two-dimensional matrix $A={A(a,b)}$ is said to be \emph{totally
monotone} with respect to row maxima if the following relation
holds:
\begin{equation}\label{eqn:total_mono}
\begin{split}
A(a,b)\leq A(a,b')\Rightarrow A(a',b)\leq A(a',b'),\ a<a',b<b'.
\end{split}
\end{equation}
A sufficient condition for \eref{eqn:total_mono} is
\begin{equation}\label{eqn:Monge}
A(a,b')+ A(a',b)\leq A(a,b)+ A(a',b'),\ a<a',b<b'
\end{equation} which is also known as the Monge condition.
If an $n\times n$ matrix $A$ is totally monotone, then the row
maxima of $A$ can be found in $O(n)$ time \cite{aggarwal}.

To apply the linear-time matrix search algorithm to the joint source-channel
MDSQ decoding problem, we only need to show that matrix $A_n$ satisfies the total
monotonicity. Substituting $A_n$ in \eref{eqn:matrix_definition} for
$A$ in \eref{eqn:Monge}, we have
\begin{equation}\label{eqn:sufficient}
\begin{split}
\log P(a|b')+\log P(a'|b)\leq \log P(a|b) +\log P(a'|b'),\\ a<a',
b<b'
\end{split}
\end{equation}
which is a sufficient condition for $A_n$ to have the total
monotonicity and therefore, for the fast algorithm to be applicable.
This condition, which depends only on the source statistics not the
channels, is exactly the same as the one derived in~\cite{Wu}.  It
was shown by \cite{Wu} that \eref{eqn:sufficient} holds if the
source is Gaussian Markovian, which includes a large family of
signals studied in practice and theory.

Finally, we conclude that the time complexity of MAP decoding of
MDSQ can be reduced to $O(L N K)$ for Gaussian Markov sequences.

\bibliographystyle{IEEEtran}
\bibliography{JSC_MD}

\end{document}